\documentclass{sig-alternate-2013}

\usepackage{graphicx}
\usepackage{caption}
\usepackage{subfig}
\usepackage{hyperref}

\usepackage{color}
\usepackage{lmodern}
\usepackage[T1]{fontenc}

\DeclareCaptionType{copyrightbox}

\newcommand{\from}[3]{}

\def\invsa{{\vspace{-0.15in}}}

\def\userlog{{\textsf{user\_log}}}
\def\editpair{{\textsf{edit\_pair}}}
\def\raw{{\textsf{UMDWikipedia}}}
\def\VEWS{{\textsf{VEWS}}}
\newcommand{\nop}[1]{{}}

\newfont{\mycrnotice}{ptmr8t at 7pt}
\newfont{\myconfname}{ptmri8t at 7pt}
\let\crnotice\mycrnotice%
\let\confname\myconfname%

\permission{Permission to make digital or hard copies of all or part of this work for personal or classroom use is granted without fee provided that copies are not made or distributed for profit or commercial advantage and that copies bear this notice and the full citation on the first page. Copyrights for components of this work owned by others than ACM must be honored. Abstracting with credit is permitted. To copy otherwise, or republish, to post on servers or to redistribute to lists, requires prior specific permission and/or a fee. Request permissions from Permissions@acm.org.}
\conferenceinfo{KDD'15,}{August 11-14, 2015, Sydney, NSW, Australia.} 
\copyrightetc{\copyright~2015 ACM. ISBN \the\acmcopyr}
\crdata{978-1-4503-3664-2/15/08\ ...\$15.00.\\
DOI: http://dx.doi.org/10.1145/2783258.2783367}

\begin{document}
%

\title{VEWS: A Wikipedia Vandal Early Warning System}

%
%
%
%
%

\numberofauthors{3} 
%
\author{
%
%
\alignauthor
Srijan Kumar\\
       \affaddr{Computer Science Dep.}\\
       \affaddr{University of Maryland}\\
       \affaddr{College Park, 20742 MD, USA}\\
       \email{srijan@cs.umd.edu}
\alignauthor
Francesca Spezzano\\
       \affaddr{UMIACS}\\
       \affaddr{University of Maryland}\\
       \affaddr{College Park, 20742 MD, USA}\\
       \email{spezzano@umiacs.umd.edu}
\alignauthor V.S. Subrahmanian\\
       \affaddr{Computer Science Dep.}\\
       \affaddr{University of Maryland}\\
       \affaddr{College Park, 20742 MD, USA}\\
       \email{vs@cs.umd.edu}
}

\maketitle
\begin{abstract}
We study the problem of detecting vandals on Wikipedia \emph{before} any human or known vandalism detection system reports flagging potential vandals so that such users can be presented early to Wikipedia administrators. We leverage multiple classical ML approaches, but develop 3 novel sets of features. Our Wikipedia Vandal Behavior (WVB) approach uses a novel set of user editing patterns as features to classify some users as vandals.  Our Wikipedia Transition Probability Matrix (WTPM) approach uses a set of features derived from a transition probability matrix and then reduces it via a neural net auto-encoder to classify some users as vandals. The \VEWS\ approach merges the previous two approaches. Without using any information (e.g. reverts) provided by other users,
these algorithms each have over 85\% classification accuracy. Moreover, when temporal recency is considered, accuracy goes to almost 90\%.  We carry out detailed experiments on a new data set we have created consisting of about 33K Wikipedia users (including both a black list and a white list of editors) and containing 770K edits.   We describe specific behaviors that distinguish between vandals and non-vandals. We show that \VEWS\ beats ClueBot NG and STiki, the best known algorithms today for vandalism detection. Moreover, \VEWS\ detects far more vandals than ClueBot NG and on average, detects them 2.39 edits before ClueBot NG when both detect the vandal. However, we show that the combination of \VEWS\ and ClueBot NG can give a fully automated vandal early warning system with even higher accuracy.
\end{abstract}



\newpage
\section{Introduction}
With over 4.6M articles, 34M pages, 23M users, and 134K active users, English Wikipedia is one of the world's biggest information sources, disseminating information on virtually every topic on earth. Versions of Wikipedia in other languages further extend its reach. Yet, Wikipedia is compromised by a relatively small number of vandals --- individuals who carry out acts of vandalism that Wikipedia defines as ``\emph{any addition, removal, or change of content, in a deliberate attempt to compromise the integrity of Wikipedia}" \cite{vandalism}.
Vandalism is not limited to Wikipedia itself, but is widespread in most social networks. Instances of vandalism have been reported in Facebook (vandalism of Martin Luther King, Jr.'s fan page in Jan 2011), WikiMapia and OpenStreetMaps~\cite{openstreetmap}. 

There has been considerable work on identifying vandalized pages in Wikipedia.  For instance, 
ClueBot NG~\cite{cluebot}, STiki~\cite{stiki}, and Snuggle~\cite{snuggle} use heuristic rules and machine learning algorithms to flag acts of vandalism. There is also linguistic work on finding suspicious edits by analyzing edit content
\cite{Velasco10,WestKL10,AdlerAP10,AdlerAMRW11,Potthast}. Most of these works use linguistic features to detect vandalism.

Our goal in this paper is the \emph{early identification} of vandals before any human or known vandalism detection system reports  vandalism so that they can be brought to the attention of Wikipedia administrators. This goes hand-in-hand with human reporting of vandals. But this information is not used in any of our three algorithms.\footnote{\small Just for completeness, Section ~\ref{sec:diff-with-reverts} reports on differences between vandals and benign users when reverts are considered. Our experiments actually show that using human or known vandalism detection system generated reversion information improves the accuracy of our approaches by only about 2\%, but as our goal is early detection, \VEWS\ ignores reversion information.}

This paper contains five main contributions.

1. We define a novel set of ``behavioral features'' that capture edit behavior of Wikipedia users.

2. We conduct a study showing the differences in behavioral features for vandals vs. benign users.

3. We propose three sets of features that use no human or known vandal detection system's reports of vandalism to predict which users are vandals and which ones are benign. These approaches use the behavioral features from above and have over 85\% accuracy.  Moreover, when we do a classification using data from previous $n$ months up to the current month, we get almost 90\% accuracy. We show that our \VEWS\ algorithm handily beats today's leaders in vandalism detection - ClueBot NG (71.4\% accuracy) and STiki (74\% accuracy). Nonetheless, \VEWS\ benefits from ClueBot NG and STiki - combining all three gives the best predictions.

4. \VEWS\ is very effective in early identification of vandals. \VEWS\ detects far more vandals (15,203) than ClueBot NG (12,576).
On average, \VEWS\ predicts a vandal after it makes (on average) 2.13 edits, while ClueBot NG needs 3.78 edits. Overall, the combination of \VEWS\ and ClueBot NG gives a fully automated system without any human input to detect vandals (STiki has human input, so it is not fully automated).

5. We develop the unique \raw\ data set that consists of about 33K users, about half of whom are on a white list, and half of whom are on a black list.

\vspace*{-3mm}
\section{Related Work}\label{sec:relwork}
To date, almost all work on Wikipedia vandals has focused on the problem of identifying pages whose text has been vandalized.
The first attempt to solve this problem came directly from the Wikipedia community with the development of \emph{bots} implementing simple heuristics and machine learning algorithms to automatically detect page vandalism (some examples are ClueBot NG~\cite{cluebot} and STiki~\cite{stiki}).

The tools currently being used to detect vandalism on Wikipedia are ClueBot NG and STiki. 
ClueBot NG is the state-of-the-art bot being used in Wikipedia to fight vandalism. It uses an artificial neural network to score edits and reverts the worst-scoring edits.
STiki \cite{stiki} is another tool to help trusted users to revert vandalism edits using edit metadata (editor's timestamp, user info, article and comment), user reputation score and textual features. STiki leverages the spatio-temporal properties
of edit metadata to assign scores to each edit, and uses human or bot reverted edits of the user to incrementally maintain a user reputation score \cite{WestKL10}.
In our  experiments, we show that our method beats both these tools in finding vandals.

A number of approaches such as \cite{Velasco10,WestKL10,AdlerAP10,AdlerAMRW11,missing} (see \cite{ferschke2013survey} for a survey) use feature extraction (including some linguistic features) and machine learning and validate them on the PAN-WVC-10 corpus: a set of 32K edits annotated by humans on Amazon Mechanical Turk.
\cite{AdlerAP10} builds a classifier by using the features computed by WikiTrust \cite{wikitrust} which monitors edit quality, content reputation, and content-based author reputation.\footnote{\small
WikiTrust cannot be used to detect vandals immediately, as it requires a few edits made on the same article to judge an edit and modify the user reputation score. WikiTrust was discontinued as a tool to detect vandalism in 2012 due to poor accuracy and unreliability.}
By combining all the features (NLP, reputation and metadata) from \cite{Velasco10,AdlerAP10} and STiki tool \cite{WestKL10}, it is possible to obtain a classifier with better accuracy~\cite{AdlerAMRW11}. 

Past efforts differ from ours in at least one of the two respects: they i)  predict whether an edit is vandalism, and not whether a user is a vandal, or ii) take into account factors that involve human input (such as number of user's edits reverted). We have not used textual features at all (and therefore, we do not rely on algorithms/heuristics that predict vandalism edits). However, we show that the combination of linguistic (from ClueBot NG and STiki) and non-linguistic features (from \VEWS\ algorithm) gives the best classification results. Moreover, we show that a fully automated (without human input) effective vandal detection system can be created by combining \VEWS\ and ClueBot NG.

Our work is closer in spirit to \cite{WestL12} which studies how humans navigate through Wikipedia in search of information.
They 
proposed an algorithm to predict the user's intended target page, given the click log. In contrast, we study users' edit patterns and differentiate between users based on the pages he/she has edited. Other studies look at users' web navigation and surfing behavior \cite{cockburn2001web,catledge1995characterizing} and why users re-visit certain pages \cite{adar}.
By using patterns in edit histories and egocentric network properties, \cite{Welser11} proposes a method to identify the social roles played by Wikipedia users (substantive experts, technical editors, vandal fighters, and social networkers), but don't identify vandals.

\section{The UMDWikipedia Dataset}
We now describe the \raw\ dataset\footnote{\small The data set is available at \url{http://www.cs.umd.edu/~vs/vews}} which captures various aspects of the edits made by 
both vandals and benign users.\footnote{\small We only studied users with registered user names. } The \raw\ dataset consists of the following components.

\textbf{Black list DB.} This consists of all 17,027 users that registered and were blocked by Wikipedia administrators for vandalism between January 01, 2013 and July 31, 2014. We refer to these users as \textit{vandals}. 

\textbf{White list DB.} This is a randomly selected list of 16,549 (benign) users who registered between January 01, 2013 and July 31, 2014 and who are not in the black list. 

\textbf{Edit Meta-data DB.} This database is constructed using the Wikipedia API \cite{wikiApi} and 
has the schema 
\vspace{-2mm}
$$(User,Page,Title,Time,Categories,M)$$ 

\vspace{-2mm}
A record of the form $(u,p,t,t',C,m)$ says that at time $t'$, user $u$ edited the page $p$ (which is of type $m$ where $m$ is either a normal page or a meta-page\footnote{\small Wikipedia pages can either be normal article pages or can be discussion or ``talk'' pages where users may talk to each other and discuss edits.}),
which has title $t$ and has list $C$ of Wikipedia categories attached to it.\footnote{\small Note that Wikipedia assigns a category to each article from a category tree --- this therefore labels each page with the set of categories to which it belongs.}
All in all, we have 770,040 edits: 160,651 made by vandals and 609,389 made by benign users.

\begin{table*}
\footnotesize
\begin{tabular}{|l|}
\hline
Whether $p_2$ is a meta-page or normal page.\\\hline
Time difference between the two edits: less than 3 minutes (very fast edit), less than 15 minutes (fast edit),\\more than 15 minutes (slow edit).\\\hline
Whether or not $p_2$ is the first page ever edited by the user.\\\hline
Whether or not $p_2$ is a page that has already been edited by the user before ($p_2$ is a re-edit) and, if yes\\
\ \ \ -  Whether or not $p_1$ is equal to $p_2$ (i.e. were two consecutive edits by the same user applied to the same page);\\
\ \ \ -  Whether of not a previous edit of $p_2$ by the user $u$ has been reverted by any other Wikipedia user.\\
Otherwise, $p_2$ is a page edited for the first time by user $u$. In this case, we include the following data:\\
\ \ \ -  the minimum number of links from $p_1$ and $p_2$ in the Wikipedia hyper-link graph: more than 3 hops, at most 3 hops,\\
\ \ \ \ \ or not reachable;\\
\ \ \ -  the number of categories $p_1$ and $p_2$ have in common: none, at least one, or $null$ if category information is not available.\\\hline
\end{tabular}
\caption{Features used in the \editpair\ and \userlog\ datasets to describe a consecutive edit $(p_1,p_2)$ made by user $u$.}\label{tab:editPairFeatures}
\vspace{-4mm}
\end{table*}

\textbf{Edited article hop DB.} This database specifies, for each pair $(p_1,p_2)$ of pages that a user consecutively edited, the minimum distance in the Wikipedia hyper-link graph\footnote{\small This is the graph whose vertices are pages and where there is an edge from page $p_1$ to $p_2$ if $p_1$ contains a hyper-link to $p_2$.} from $p_1$ to $p_2$. We used the code provided by \cite{hopsDB}.

\textbf{Revert DB.}  Just for the one experiment we do at the very end, we use the edit
reversion dataset provided by \cite{revertDB} which marks an edit  ``reverted" if it has been reverted within next 15 edits on the page. \cite{Kittur} suggests that 94\% of the reverts are detected by the method used to create the dataset. Therefore, we use this dataset as ground truth to know whether the edit was reverted or not. 
\emph{Note that this information is not needed as a feature in our algorithm for prediction, but to analyze the property of reversion across vandals and benign users.}
Observe that Revert DB also contains the information whether the reversion has been made by ClueBot NG. We use this to compare with ClueBot NG.

\textbf{STiki DB.} We use the STiki API~\cite{stikiAPI} to collect STiki vandalism scores, and the raw 
features used to derive these scores (including the user reputation score). We use vandalism and user scores \emph{only} to compare with STiki.

\invsa \smallskip\paragraph{Edit Pair and User Log Datasets} To analyze the properties of edits made by vandals and benign users, we create two additional datasets using the data in the \raw\ dataset.

\textbf{Edit Pair Dataset.} The \editpair\ dataset contains a row for each edit $(u,p_1,p_2,t)$, where $u$ is a user id,  $(p_1,p_2)$ is a pair of Wikipedia pages that are consecutively edited by user $u$, and $t$ is the time stamp of the edit made on $p_2$.  Note that $p_1$ and $p_2$ could be the same if the user makes two edits, one after another, on the same page. Each row contains the values of the features shown in 
Table~\ref{tab:editPairFeatures} computed for the edit $(u,p_1,p_2,t)$. These features describe the properties of page $p_2$ with respect to page $p_1$. 

\textbf{User Log Dataset.}
The chronological sequence of each consecutive pair $(p_1,p_2)$ of pages edited by the same user $u$ corresponds to a row in this dataset.
Each pair $(p_1,p_2)$ is described by using the features from Table~\ref{tab:editPairFeatures}.
This dataset is derived from the \editpair\ dataset. 
It captures a host of temporal information about each user, suggesting how he/she navigated through Wikipedia and the speed with which this was done.

\begin{figure*}[!ht]
        \centering
        \subfloat[]{
                \includegraphics[trim = 0mm 0mm 1.5cm 0mm, clip, width=0.30\textwidth]{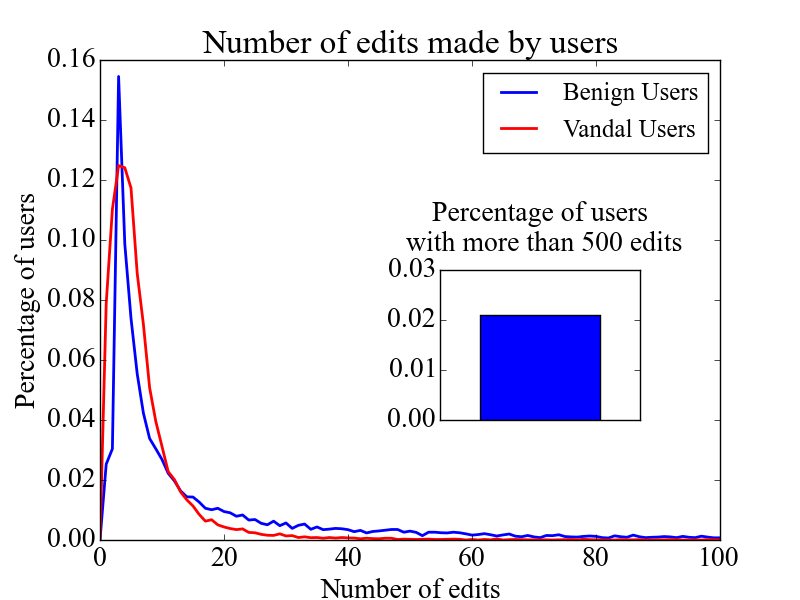}\label{fig:histo1}
        }%
        ~ 
        \subfloat[]{
                \includegraphics[trim = 0mm 0mm 15mm 0mm, clip, width=0.30\textwidth]{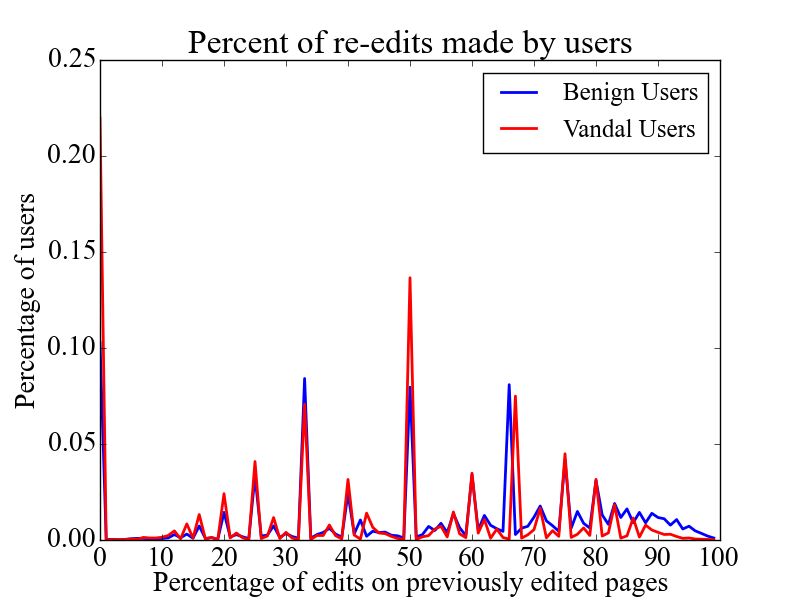}\label{fig:histo4}
        }
        ~
        \subfloat[]{
                \includegraphics[trim = 0mm 0mm 15mm 0mm, clip, width=0.30\textwidth]{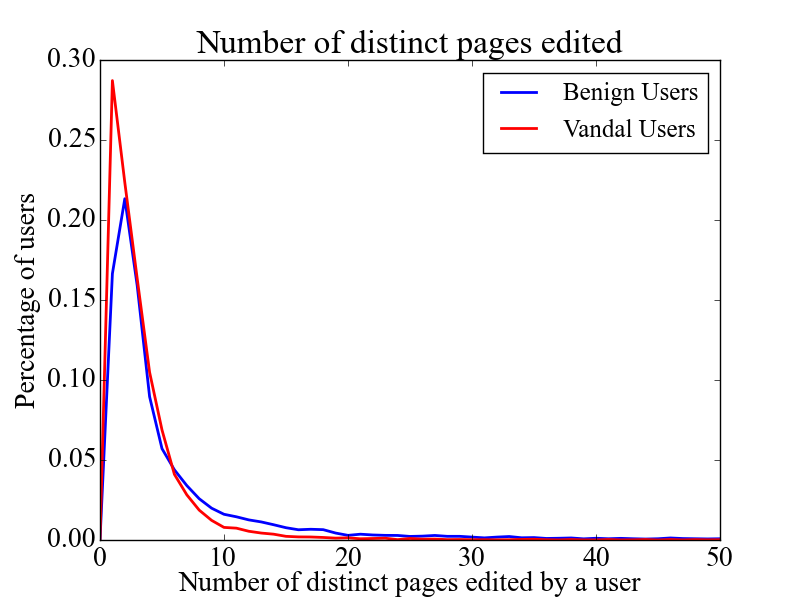}\label{fig:histo3}
        }%
          
        \vspace{-4mm}
        \subfloat[]{
                \includegraphics[trim = 2mm 0mm 15mm 5mm, clip, width=0.30\textwidth]{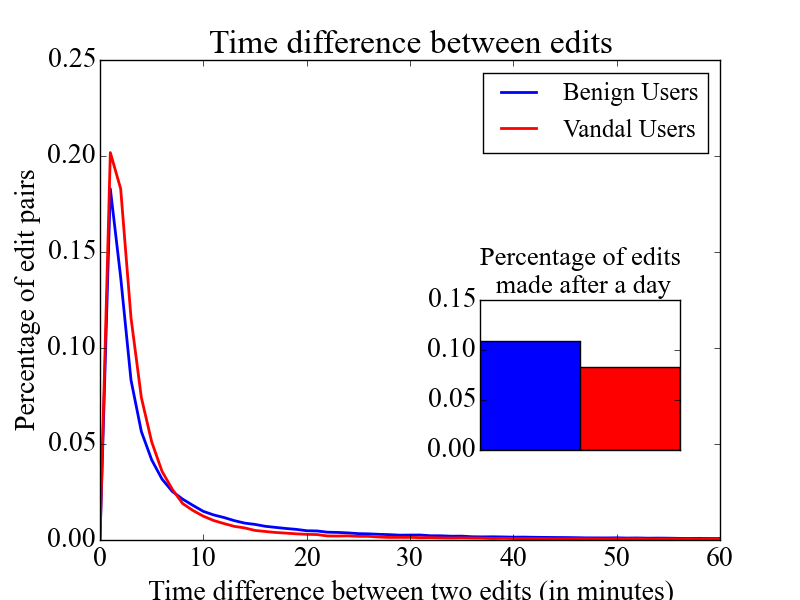}\label{fig:histo2}
       }
        \subfloat[]{
                \includegraphics[trim = 5mm 1mm 15mm 5mm, clip, width=0.30\textwidth]{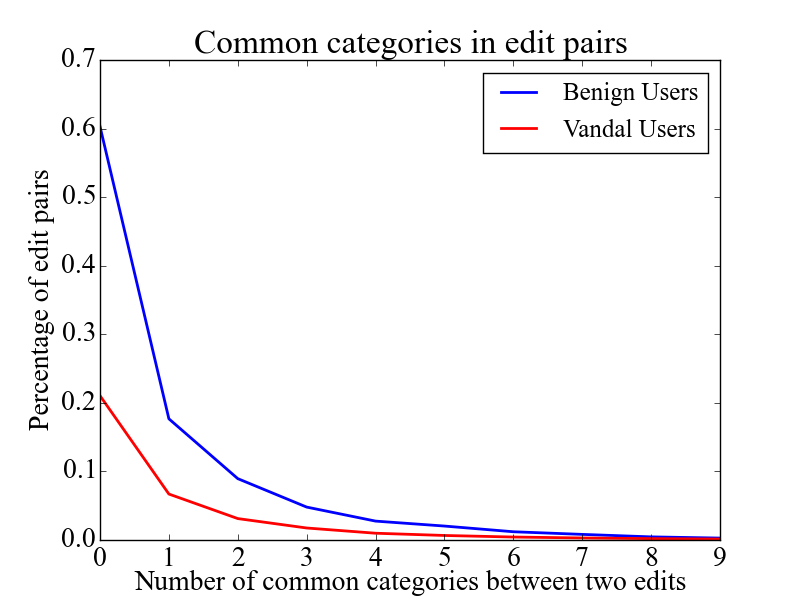}\label{fig:histo5}
        }%
        ~ 
        \subfloat[]{
                \includegraphics[trim = 5mm 0mm 15mm 4mm, clip, width=0.30\textwidth]{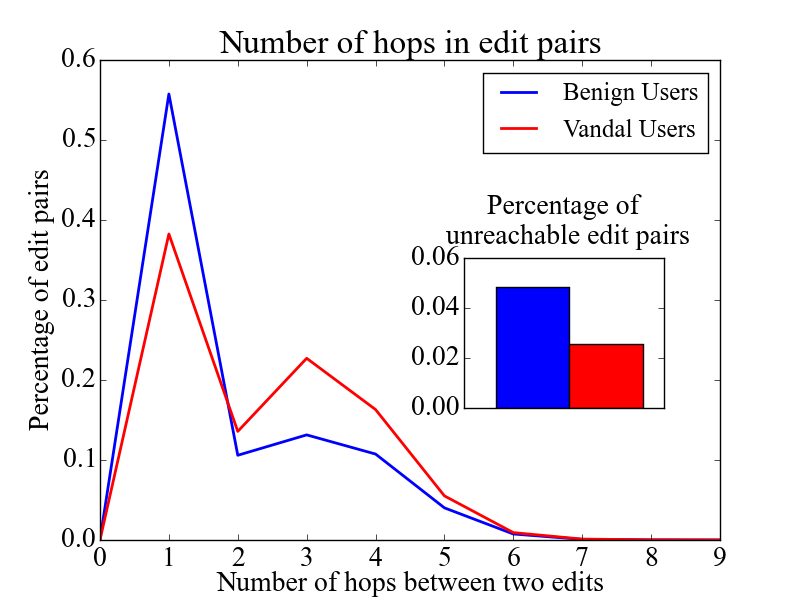}\label{fig:histo6}
        }
        \caption{\small Plots showing the distribution of different properties for \raw\ and \editpair\ datasets.}\label{fig:histograms}
        \vspace{-4mm}
\end{figure*}

\vspace*{-0.2cm}
\section{Vandal vs. Benign User Behaviors}
In this section, we statistically analyze editing behaviors of vandals and benign users in order to identify behavioral similarities and differences. 

Figure~\ref{fig:histograms} shows the distributions of different properties that are observed in the \editpair\ dataset. 
Figures~\ref{fig:histo1}-~\ref{fig:histo3} show the percentage of users on the $y$-axis as we vary the number of edits, number of distinct pages edited and the percentage of re-edits on the $x$-axis. These three graphs show near identical behavior. 

Figures~\ref{fig:histo2}-~\ref{fig:histo6} show the percentage of edit pairs $(u,p_1,p_2)$ on the $y$-axis as we vary time between edits, number of common categories between edited pages $p_1$ and $p_2$ and  number of hops between $p_1$ and $p_2$. The behavior of users in terms of time taken between edits is nearly identical. The last two graphs show somewhat different behaviors between vandals and benign users. Figure~\ref{fig:histo5} shows that the percentage of edit pairs involving just one, two, or three common categories is 2-3 times higher for benign users than for vandals.
Likewise, figure~\ref{fig:histo6} shows that for benign users, the percentage of edit pairs involving exactly one hop is 1.5 times that of vandals, but the percentage of edit pairs involving 3-4 hops is much higher for vandals than for benign users.

In all the histograms in Figure~\ref{fig:histograms}, the null hypothesis that the distribution for vandals and benign users have identical average has p-value > 0.05.
As this fails to say that their behavior is not similar, we do a more in-depth analysis to  distinguish between them.
So, we perform a frequent itemset mining step on the \editpair\ and \userlog\ datasets.
Figure~\ref{fig:plots} summarizes the results.

\subsection{Similarities between Vandal and Benign User Behavior (w/o reversion features)}
\begin{figure*}[!ht]
        \centering
        \subfloat[\small Type of edit and time data overall edits ($\Delta t\_c$ is the elapsed time between edits on the same page).]{
                \includegraphics[width=0.36\textwidth]{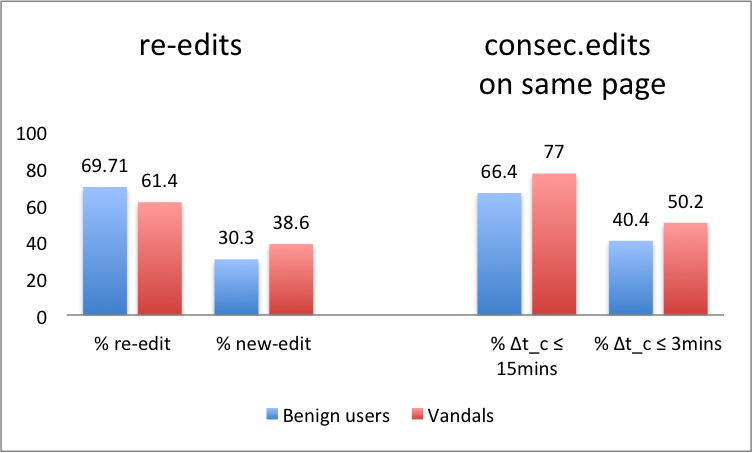}\label{fig:plot1}
        }
        ~ 
        \subfloat[\small For a consecutive edit $(p_1,p_2)$, how users surf from $p_1$ to the edit of a new page $p_2$.]{
                \includegraphics[width=0.36\textwidth]{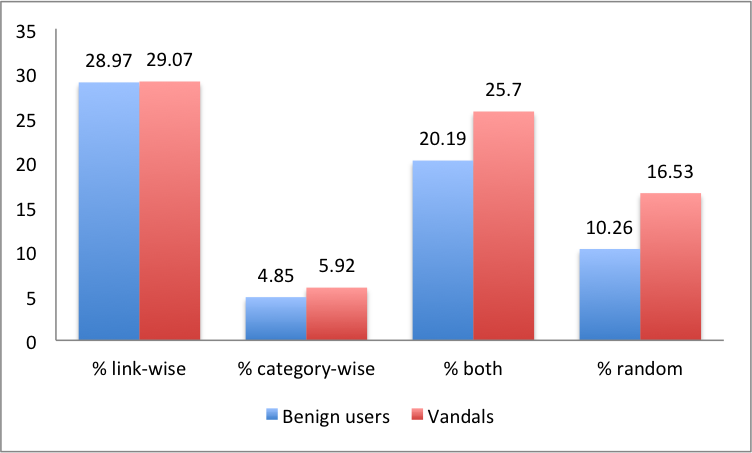}\label{fig:plot2}
        }
        
        \subfloat[\small First 4 edits in users log.]{
                \includegraphics[width=0.3625\textwidth]{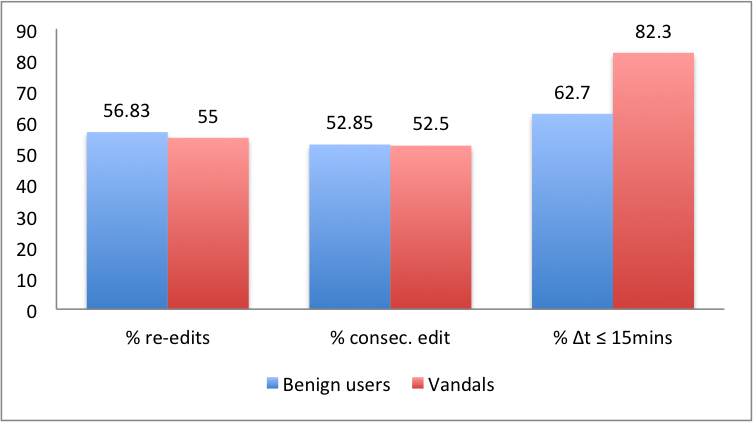}\label{fig:plot3}
        }
        ~
         \subfloat[\small Edit time.]{
                \includegraphics[width=0.355\textwidth]{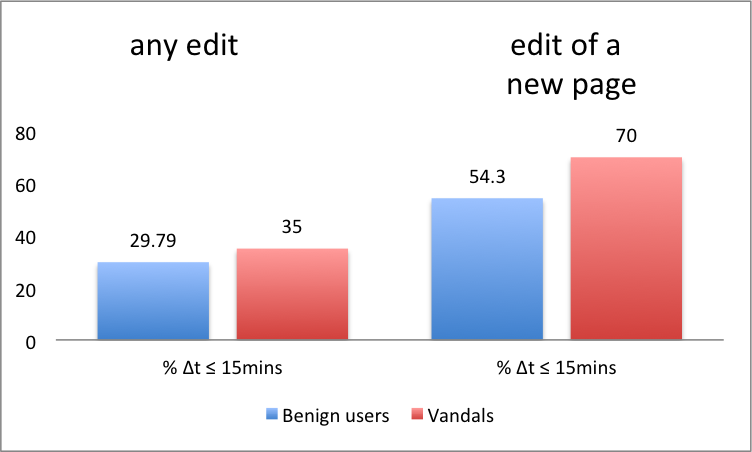}\label{fig:plot5}
        }
        
        \subfloat[\small Meta-pages editing.]{
                \includegraphics[width=0.36\textwidth]{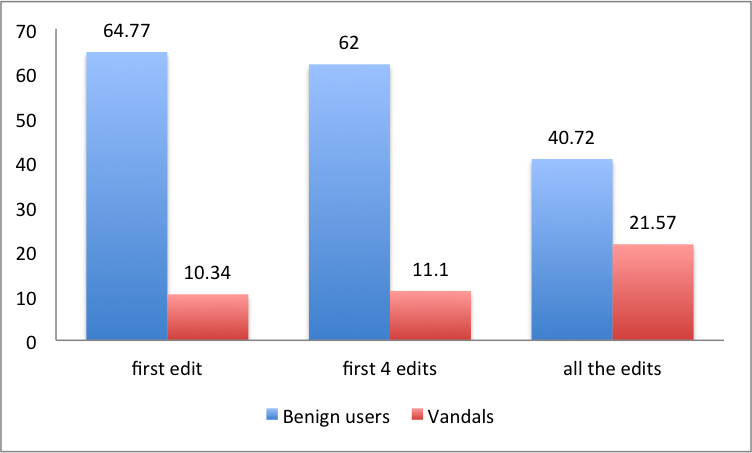}\label{fig:plot4}
        }
        ~
        \subfloat[\small Edit war (involves reversion).]{
                \includegraphics[width=0.36\textwidth]{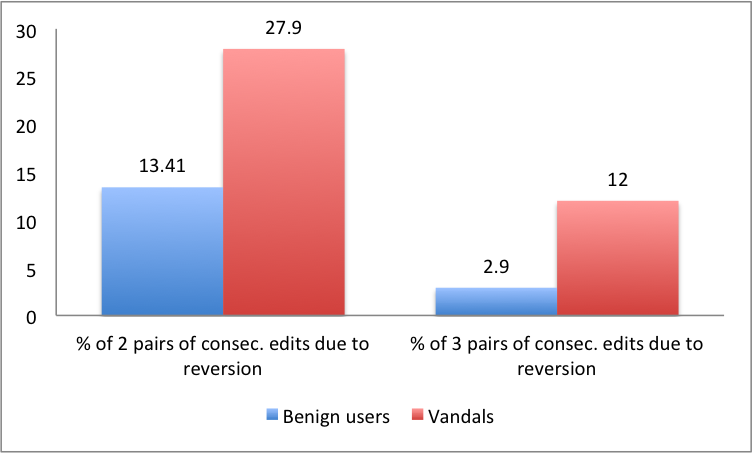}\label{fig:plot6}
        }
        \caption{\small Analogies and differences between benign users and vandals.}\label{fig:plots}
        \vspace{-4mm}
\end{figure*}

Figures~\ref{fig:plot1}, \ref{fig:plot2}, and \ref{fig:plot3} show similarities between vandal and benign user behaviors.

$\bullet$ \emph{Both vandals and benign users are much more likely to re-edit a page compared to editing a new page.} We see from 
 Figure~\ref{fig:plot1} that for vandals, the likelihood of a re-edit is 61.4\% compared to a new edit (38.6\%). Likewise,  for benign users, the likelihood of a re-edit is 69.71\% compared to a new edit (30.3\%).

$\bullet$ \emph{Both vandals and benign users consecutively edit the same page quickly.} The two rightmost bars in  Figure~\ref{fig:plot1}  show that
both vandals and benign users edit the same page fast. 77\% of such edit pairs (for vandals) occur within 15 minutes --
this number is 66.4\% for benign users. In fact, over 50\% of these edits occur within 3 minutes for vandals - the corresponding number for benign users is just over 40\%.

$\bullet$\emph{Both vandals and benign users exhibit similar navigation patterns.} 29\% of successively edited pages (for both vandals and benign users) are by following links only (no common category and reachable by hyperlinks), about 5\% due to commonality in categories only between the successively edited pages (at least one common category and not reachable by hyperlinks), and 20-25\% with commonality in both properties and linked. This is shown in 
Figure~\ref{fig:plot2}.

$\bullet$ \emph{In their first few edits, both vandals and benign users have similar editing behavior}: Figure~\ref{fig:plot3} shows just the first 4 edits made by both vandals and benign users. We see here that the percentage of re-edits and consecutive edits are almost the same in both cases.

\subsection{Differences between Vandals and Benign User Behavior (w/o reversion features)}
We also identify several behaviors which differentiate between vandals and benign users.

$\bullet$ \emph{Vandals make faster edits than benign users.} On average, vandals make 35\% of their edits within 15 minutes of the previous edit while benign users make 29.79\% of their edits within 15 minutes (Figure~\ref{fig:plot5}). 
This difference is statistically significant with a p-value of $8.2\times10^{-82}$.

$\bullet$\emph{Benign users spend more time editing a new (to them) page than vandals.} Vandals make 70\% of their edits to a page they have not edited before within 15 minutes of their last edit, while for benign users the number is 54.3\% (Figure~\ref{fig:plot5}).  This may be because a benign user must absorb the content of a new page before making thoughtful edits, while a vandal knows what he wants to say in advance and just goes ahead and says it. 

$\bullet$ \emph{The probability that benign users edit a meta-page is much higher than the same probability in the case of vandals.}
 Figure~\ref{fig:plot4}  shows that even in their very first edit, benign users have a 64.77\% chance of editing a meta-page, while this is just 10.34\% for vandals. If we look at the first 4 edits,  the percentage of edits that are on meta-pages is 62\% for benign users and just 11.1\% for vandals. And if we look at all the edits, 40.72\% of edits by normal users are on meta-pages, while only 21.57\% of edits by vandals are on meta-pages.

 \subsection{Differences between Vandals and Benign User Behavior (including reversion)}\label{sec:diff-with-reverts}
For the sake of completeness, we also analyze the data looking for differences between vandal and benign user behavior when reverts are considered --- however these differences are not considered in our vandal prediction methods.

$\bullet$ \emph{Vandals make more edits driven by reversion than benign users.}
Whenever a vandal $u$ re-edits a page $p$, in 34.36\% of the cases, $u$'s previous edit on $p$ was reverted by others.
This almost never occurs in the case of benign users --- the probability is just 4.8\%. This suggests that benign users are much more accepting of reversions than vandals.

$\bullet$\emph{The probability that a re-edit by a benign user of a page is accepted, even if previous edits by him on the same page were reverted, is much higher than for vandals}. Consider the case when a user edits a page \textit{p} after some of his prior edits on \textit{p} were reverted by other.  If the user \textit{u} is a benign user, it is more likely that his last edit is accepted. This suggests that the sequence of edits made by \textit{u} were collaboratively edited by others with the last one surviving, suggesting that \textit{u}'s reverts were constructive and were part of a genuine collaboration. Among the cases when \textit{u} re-edits a page after one of his previous edits on \textit{p} has been reverted, 89.87\% of these re-edits survive for benign users, while this number is only 32.2\% for vandals.

$\bullet$ \emph{Vandals involve themselves in edit wars much more frequently than benign users}. A user $u$ is said to participate in an edit war if there is a consecutive sequence of edits by $u$ on the same page which is reverted at least two or three times (we consider both cases). Figure~\ref{fig:plot6} shows that 27.9\% of vandals make two pairs of consecutive edits because their previous edit was reverted, but only 13.41\% of
benign users do so.  12\% of vandals make three such pairs of consecutive edits, compared to 2.9\% in the case of benign users.

$\bullet$ \emph{The probability that benign users discuss their edits is much higher than the probability of vandals doing so}. In 31.3\% of the cases when a benign user consecutively edits a page $p$ twice (i.e. the user is actively editing a page), he then edits a meta page. With vandals, this probability is 11.63\%. This suggests that benign editors discuss edits on a meta-page after an edit, but vandals do not (perhaps because doing so would draw attention to the vandalism). In addition there is a 24.41\% probability that benign users will re-edit a normal Wikipedia page after editing a meta-page while this happens much less frequently for vandals (only 6.17\% vandals do such edits). This indicates that benign users, after discussing relevant issues on meta pages, edit a normal Wikipedia page.

$\bullet$ \emph{Benign users consecutively surface edit pages a lot}. 
We define a surface edit by a user $u$ on page $p$ as: i) a consecutively edit on the same page $p$ twice by $u$, and ii) the edit is not triggered by $u$'s previous edit on $p$ being reverted, and iii) made within 3 minutes of the previous edit by $u$.  50.94\% benign users make at least one surface edit on a meta page, while only 8.54\% vandals do so. On normal pages, both benign and normal users make such edits - 
there are 37.94\% such cases for benign users and 36.94\% for vandals. Over all pages, 24.24\%  benign users make at least 3 consecutive surface edits not driven by reversion, but only 7.82\%  vandals do so.

In conclusion: \textsf{(i)} 
Vandals make edits at a faster rate than benign users.
\textsf{(ii)} Vandals are much less engaged in edits of meta pages, i.e. they are less involved in discussions with the community.

\vspace*{-4mm}
\section{Vandal Prediction}
In the following sections, we use the insights from the previous section to classify vandals and benign users.
Our vandal prediction methods use multiple known classifiers (SVM, decision trees, random forest and k-nearest neighbors) with different sets of features.
In the accuracies reported in this section, the results are computed with SVM, as it gives the highest accuracy as reported in Section 6 using a 10-fold cross validation.
\emph{All  features used for vandal prediction are behavior based and include no human generated revert information whatsoever. Thus, these approaches form an early warning system for Wikipedia administrators.}

\newpage
\subsection{Wikipedia Vandal Behavior (WVB) Approach}
WVB uses features derived from consecutive edits. These are found by frequent pattern mining of the \userlog\ dataset. Specifically, we extract the frequent patterns on both benign and vandal user logs -- then, for each frequent pattern of benign users, we compute the frequency of the same pattern for vandals and vice versa. Finally, we select the patterns having significant frequency difference between the two classes. The resulting features are described below:

1. \textbf{Consecutive re-edit, slowly (crs)}: whether or not the user edited the same page consecutively with a time gap exceeding 15 minutes.

2. \textbf{Consecutive re-edit, very fast (crv)}: whether or not the user edited the same page consecutively and less than 3 minutes passed between the two edits.

3. \textbf{Consecutive re-edit of a meta-page (crm)}: the number of times the user re-edited the same meta-page, consecutively.

4. \textbf{Consecutive re-edit of a non-meta-page (crn)}: whether or not the user re-edited the same non-meta-page, consecutively.

5. \textbf{Consecutive re-edit of a meta-page, very fast (crmv)}: whether or not the user re-edited the same meta-page, consecutively, and less than 3 minutes passed between the two edits.

6. \textbf{Consecutive re-edit of a meta-page, fast (crmf)}: whether or not the user re-edited the same meta-page, consecutively, and 3 to 15 minutes passed between the two edits.

7. \textbf{Consecutive re-edit of a meta-page, slowly (crms)}: whether or not the user re-edited the same meta-page, consecutively, and more than 15 minutes passed between the two edits.

8. \textbf{Consecutively re-edit fast and consecutively re-edit very fast (crf\_crv)}: whether or not the following pattern is observed in the user log - the user re-edited the same article within 15 minutes, and later re-edited a (possibly different) article and less than 3 minutes passed between the second pair of edits.

9. \textbf{First edit meta-page (fm)}: whether or not the first edit of the user was on a meta-page.
This in itself is quite a distinguishing feature, because usually vandals first edit a non-meta page and benign users first edit a meta-page. Therefore, this becomes quite an important feature for distinguishing the two.

10. \textbf{Edit of a new page at distance at most 3 hops, slowly (ntus)}: 
whether or not the user edited a new page (never edited by him before) $p_2$ which is within 3 hops or less of the previous page $p_1$ that he edited and either $p_1$ or $p_2$'s category is unknown\footnote{\small This happens mostly for meta-pages though it can occasionally also happen for normal (non-meta) pages.} and the time gap between the two edits exceeds 15 minutes.

11. \textbf{Edit of a new page at distance at most 3 hops slowly and twice (nts\_nts)}: whether or not there are two occurrences of the following feature in the user log:
\emph{Edit of a new page at distance at most 3 hops, slowly (nts)}, i.e. in a pair $(p_1,p_2)$ of consecutive edits, whether or not the user edited a new page $p_2$ (i.e. never edited before) such that $p_2$ can be reached from $p_1$ link-wise with at most 3 hops, and more than 15 minutes passed between the edit of $p_1$ and $p_2$.

\smallskip
\emph{In predicting vandals, we do not use
 any feature involving human identification of vandals (e.g. number of edits and reversion)} 
because number of edits made has a bias towards benign users as they tend to perform more edits, while vandals perform fewer edits because they get blocked. Any feature that has a negative human intervention (number of reversions, number of warnings given to the user on a talk page, etc.) already indicates human recognition that a user may be a vandal. We explicitly avoid such features so that we provide Wikipedia administrators with a fully automated vandal early warning system.

\begin{figure}
\centering
\includegraphics[trim = 1cm 0.6cm 2.4cm 0mm, clip, width=7.5cm]{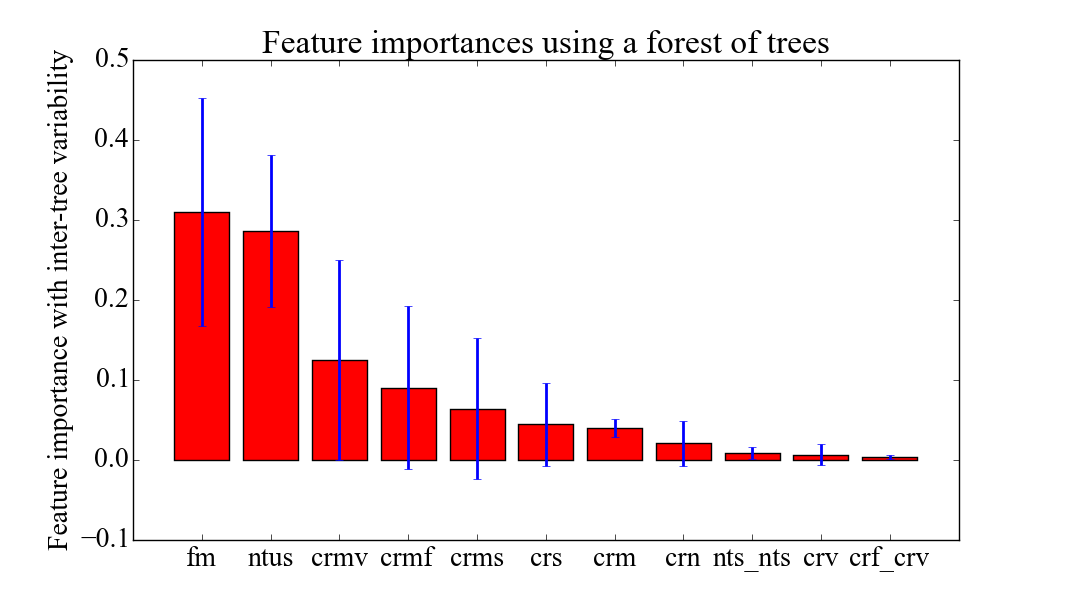}
\caption{\small Importance of features (w/o reversion).}\label{fig:importance}
\vspace{-3mm}
\end{figure}

\begin{figure}
\centering
\includegraphics[trim = 0mm 0mm 0mm 0mm, clip, width=7.5cm]{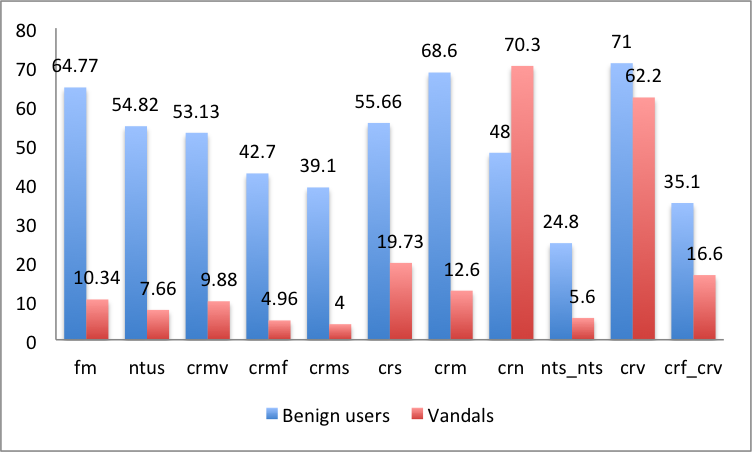}
\caption{\small Percentage of vandals and benign users with particular features (w/o reversion).}\label{fig:first3}
\vspace{-6mm}
\end{figure}

{\bf \emph{Feature importance:}}
We compute the importance of the features described above by using the fact that the depth of a feature used as a decision node in a tree captures the relative importance of that feature w.r.t.  the target variable. Features at the top of the tree contribute to the final prediction decision of a larger fraction of  inputs. The expected fraction of samples they contribute to can be used to estimate their importance.
Figure~\ref{fig:importance} shows the importance of the different features for the classification task, which was computed by using a forest of 250 randomized decision trees (extra-trees~\cite{geurts2006extremely}). The red bars in the plot show the feature importance using the whole forest, with their variability across the trees represented by the blue bars. From the figure, it is clear that the features - \emph{fm}, \emph{ntus} and \emph{crmv} - are the three most descriptive features for the classes. These are shown in greater detail in Figure~\ref{fig:first3}. Let us look into each of them one by one.

$\bullet$ \emph{If the very first page edited by user $u$ is a normal (non-meta) page, then $u$ is much more likely to be a vandal (64.77\%) than a benign user (10.34\%).} The $fm$ feature tells us that when a user's first edit is on a normal page, the user is much more likely to be a vandal.
 
$\bullet$ \emph{Benign users are likely to take longer to edit a new page than a vandal ($ntus$).}
The probability that a benign user takes more than 15 minutes to edit the next page in an edit pair $(p_1,p_2)$ when $p_2$ is within 3 hops of $p_1$, and $p_1$ or $p_2$'s category is unknown is much higher (54.82\%) than for  vandals (7.66\%). This suggests that benign users take longer to edit pages than vandals, possibly because they are careful and anxious to do a good job. Moreover, as $p_1$ or $p_2$ have no categories, the page is more likely to be a meta-page.


$\bullet$ \emph{Benign users are much more likely to re-edit the same meta-page quickly (within 3 minutes) than vandals.}
This usually happens when there is a minor mistake on the page, and the user edits to correct it. Note that this again has the feature that the edit was made on a meta page. Benign users are much more likely to make such edits (53.13\%) than vandals (9.88\%).

The top three features indicate that editing meta versus normal Wikipedia pages is a strong indicator of whether the user is benign. Intuitively, vandals vandalize heavily accessed pages  and so normal pages are their most common target. On the other hand, benign users interact and discuss issues with other users about the content of the edit, and this discussion is done on meta pages. 

{\bf \emph{Accuracy:}}
Using an SVM classifier, the WVB approach obtains an accuracy of 86.6\% in classifying Wikipedia users as vandals or benign on our entire \userlog\ dataset. 

\subsection{Wikipedia Transition Probability Matrix (WTPM) Approach}
The Wikipedia Transition Probability Matrix (WTPM)  captures the edit summary of the users.
The states in WTPM correspond to the space of possible vectors of features associated with any edit pair $(p_1,p_2)$ carried out by a user $u$. By looking at Table~\ref{tab:editPairFeatures}, we see that there are 2 options for whether $p_2$ is a meta-page or not, 3 options for the time difference between edits $(p_1,p_2)$, and so forth. This gives us a total of 60 possible states.  Example states include: \emph{consecutively re-edit a normal-page within 15 minutes} ($s_1$), or \emph{edit a new normal page $p_2$ within 3 hops from $p_1$ and no common categories within 3 minutes} ($s_2$), etc.

The transition matrix $T(u)$ of user $u$ captures the probability $T_{ij}(u)$ that user $u$ goes from state $s_i$ to $s_j$. $T_{ij} = \frac{N(s_i, s_j)}{\sum_{k}N(s_i,s_k)}$, where $N(s_i,s_j)$ is the number of times the user went from state $s_i$ to $s_j$. This gives a (usually sparse) transition matrix of size 60 $\times$ 60 = 3600. 

The intuition behind using WTPM as features for classification is that the transition probability from one state to the other for a vandal may differ from that of a benign user. Moreover, the states visited by vandals may be different from states visited by benign users (for example, it turns out that
benign users are more likely to visit a state corresponding to ``first edit on meta page", than vandals do).

We create a compact and distributed representation of $T(u)$ using an auto-encoder\cite{autoencoder}
 --- this representation provides the features for our SVM classifier.
When doing cross-validation, we train the auto-encoder using the training set with input from both benign users and vandals. We then take the value given by the hidden layer for each input as the feature for training a classifier. For predicting output for the test set, we give each test instance as input to the auto-encoder and feed its representation from the hidden layer into the classifier. Note that the auto-encoder is trained only on the training set, and the representation for the test set is derived only from this learned model.

{\bf \emph{Accuracy:}}
With a neural net auto-encoder of 400 hidden units and with SVM as the classifier, the WTPM approach gives an accuracy of 87.39\% on the entire dataset.

\subsection{VEWS Algorithm}
The \VEWS\ approach merges all the features used by both the WVB approach and the WTPM approach.
The resulting accuracy with a SVM classifier slightly improves the accuracy of classification to 87.82\%.

\section{Vandal Prediction Experiments} 
We used the popularly used machine learning library called Scikit-learn \cite{scikit} for our experiments and the deep learning library Theano \cite{theano} for training the auto-encoder. 

\textbf{Experiment 1: Overall Classification Accuracy.} Table~\ref{tab:confusion_matrix} shows the overall classification accuracy of all three approaches by doing a 10-fold cross validation using an SVM classifier, together with the true positive, true negative, false positive, and false negative rates. We see that TP and TN rates are uniformly high, and FP and FN rates are low, making SVM an excellent classifier.

We use McNemar's paired test to check whether the three approaches produce different results.
For all pairs among the three approaches, the null hypothesis that they produce the same results is rejected with p-value < 0.01, showing statistical significance.
Overall, \VEWS\ produces the best result even though it has slightly lower true positives and slightly more false negatives than WTPM.

We also classified using the \VEWS\ approach with decision tree classifier, random forest classifier (with 10 trees) and k-nearest neighbors classifier (with k = 3) which gave classification accuracy of 82.82\%, 86.62\% and 85.4\% respectively. We experimented with other classifiers as well, but they gave lower accuracy.

\begin{table}
\centering
\begin{tabular}{|c||c||c|c|c|c|}
\hline
& Accuracy & TPR & TNR & FPR & FNR\\\hline
WVB & 86.6\% & 0.85 & 0.89 & 0.11 & 0.15 \\\hline
WTPM & 87.39\% & 0.88 & 0.90 & 0.10 & 0.12\\\hline
\VEWS &  87.82\% & 0.87 & 0.92 & 0.08 & 0.13\\\hline
\end{tabular}
\caption{\small Table showing the accuracy and statistical values derived from the confusion matrix for the three approaches, on the entire dataset and averaged over 10 folds (without reversion features). The positive and negative class represent benign and vandal users, respectively.}\label{tab:confusion_matrix}
\vspace{-4mm}
\end{table}

\begin{figure}[t]
\centering
  \includegraphics[trim = 0cm 0.5mm 1.5cm 0.5cm, clip, width=6.6cm]{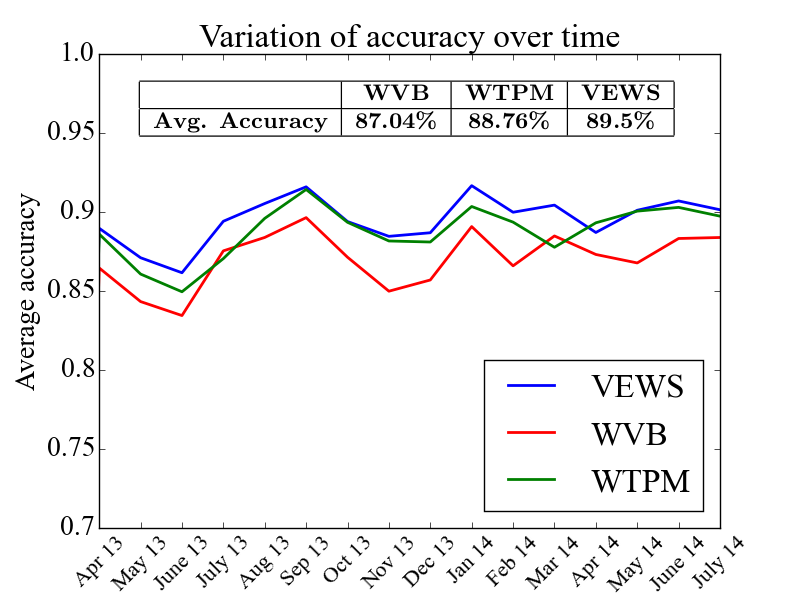}
  \caption{\small Plot showing variation of accuracy when training on edit log of users who started editing within previous 3 months (without reversion features). The table reports the average accuracy of all three approaches.}\label{fig:accuracy_time_1}
\vspace{-6mm}
\end{figure}

\textbf{Experiment 2: Improvement with Temporal Recency.}
The previous experiment's cross validation randomly selects samples from the entire dataset for training and validation. But in the real world, a vandal's behavior may be more closely related to other vandals' recent behavior. To check this, 
starting from April 2013, for each month $m$, we train our algorithms with data from all the users who started editing on Wikipedia within the previous three months, i.e. in months $m-3, m-2$ and $m-1$. $m$ is varied until July 2014. We then use the learned model to predict whether a user is vandal or benign among the users who made their first edit in month $m$. The variation of accuracy is shown in Figure~\ref{fig:accuracy_time_1}. The highest accuracy of 91.66\% is obtained with the \VEWS\ approach, when predicting for users who started editing in January 2014 and training is done with users from October-December 2013. The average accuracy for the three approaches over all the time points is also shown in Figure~\ref{fig:accuracy_time_1}.

\begin{figure}[t]
\centering
  \includegraphics[trim = 0.5cm 0mm 2cm 8mm, clip, width=6.6cm]{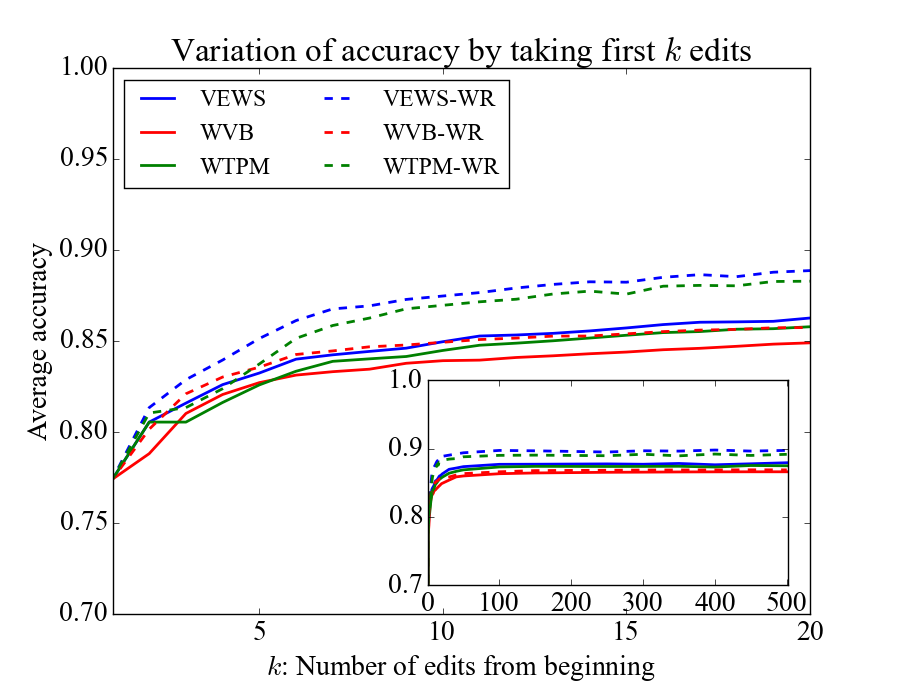}
  \caption{\small Plot showing variation of accuracy with the number of first $k$ edits. The outer plot focuses on the variation of k from 1 to 20. The inset plot shows variation of k from 1 to 500.}\label{fig:accuracy_wrev}
 \vspace{-4mm}
\end{figure}

The most important observation from  Figure~\ref{fig:accuracy_time_1} is that temporal classification accuracy for each approach is 
usually higher than the base accuracy shown in Table~\ref{tab:confusion_matrix} and
Figure~\ref{fig:accuracy_wrev} (described later in Experiment~4). We attribute this to the fact that in the previous experiment, we use cross-validation without considering temporal  information when creating the folds. This experiment, on the other hand, predicts vandals based on what is learned during the previous three months. 

Figure~\ref{fig:accuracy_time_1} shows that the approaches are consistent over time in separating vandals from benign users. At all times, the approaches have at least 85\% classification accuracy, with the exception of the case when using WVB during months May and June, 2013.

\begin{figure}[t]
\centering
  \includegraphics[trim = 0.2cm 0mm 1cm 5mm, clip, width=6.6cm]{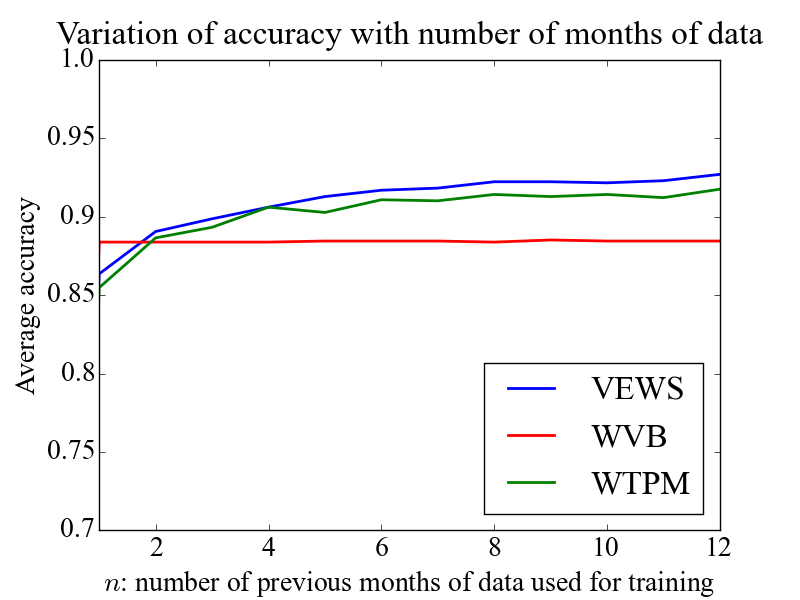}
  \caption{\small Plot showing the change in accuracy by varying the training set of users who started editing Wikipedia at most $n$ months before July 2014. The testing is done on users who started editing in July 2014.}\label{fig:accuracy_time_2}
\end{figure}

\textbf{Experiment 3: Varying Size of Training Set on Classification Accuracy.}
We designed an experiment to study the affect of varying the size of the training set, while maintaining the temporal aspect intact. So for testing on users who made their first edit in the month of July 2014, we train the classifier on edits made by users who started editing in the previous $n$ months. We vary $n$ from 1 to 12. This preserves the temporal aspect in training, similar to the previous experiment. The variation of accuracy is shown in Figure~\ref{fig:accuracy_time_2}. There are two interesting observations: i) the accuracy of WTPM and \VEWS\ increases with the number of (training) months $n$.  ii) In contrast, WVB's accuracy is hardly affected by the number of months of training data. This is because: (i) features in WVB are binary and (ii) \emph{fm}, which is the most important feature in WVB, does not vary with time.

Experiments~2 and 3 show strong temporal dependency of user behavior on prediction of vandals. This may be due to several factors: Wikipedia may change rules and policies that affect user behavior, real world events might trigger users to make similar edits and emulate similar behaviour, etc. Such behavior traits would be highlighted when observing recent edits made by newly active users.

\textbf{Experiment 4: Effect of First $k$ User Edits.}
We study the effect of the first-$k$ edits made by the user on prediction accuracy which is averaged   
 over 10 folds of the whole dataset. 
The solid lines in Figure~\ref{fig:accuracy_wrev} show the variation in accuracy when  $k$ is varied from 1 to 500. As there is little shift in classification accuracy when $k>20$, the situation for $k=1,\ldots,20$ is highlighted.
We get an average accuracy of 86.6\%  for WVB, 87.39\% for WTPM, and 87.82\% for \VEWS\, when $k$ = 500.
It is clear that the first edit itself (was the first edit made on a meta-page or not?) is a very strong classifier, with an accuracy of 77.4\%. Accuracy increases fast when $k$ is increased to 10 for all  approaches, after which it flattens out. This suggests that a user's first few edits are very significant in deciding whether he/she is vandal or not.

\noindent\textbf{\emph{Considering reversion.}} Figure~\ref{fig:accuracy_wrev} also shows that accuracy does go up by about 2\% when we allow our three algorithms to consider reversion information. 
\emph{Please note that this experiment is merely for completeness sake and our proposed algorithm does not depend on reversion at all.}
For this experiment, we added additional reversion-driven edit features to the features used by WVB, WTPM, and \VEWS\ (and we called these approaches WVB-WR, WTPM-WR, and \VEWS-WR, respectively). These features capture whether a user edited a page after his previous edit on that page was reverted. Specifically, we extend the features - \emph{crs, crv, crm, crn, crmv, crmf, crms and crf\_crv} - to now have two types of re-edits: one that is reversion driven and one that is not. Using reversion information would mean that a human or vandalism detection system has already flagged a potential vandal. In contrast, our algorithms are able to predict vandals with high accuracy even without such input.

\textbf{Comparison with State-of-the-art tools.}
Here we evaluate our work against ClueBot NG \cite{cluebot} and STiki \cite{stiki} as they are the primary tools currently used by Wikipedia to detect vandalism. We recall that these tools are designed to detect whether the content of an article has been vandalized or not, while we focus on detecting whether a user is a vandal or not. We show that \VEWS\ handily beats both ClueBot NG and Stiki in the latter task. Interestingly, when we combine \VEWS', ClueBot NG's and STiki's features, we get better accuracy than with either of them alone. All experiments are done using 10-fold cross validation and SVM as the classifier.

\begin{figure}[t]
\centering
  \includegraphics[trim = 0.5cm 0mm 0.1cm 5mm, clip, width=6.6cm]{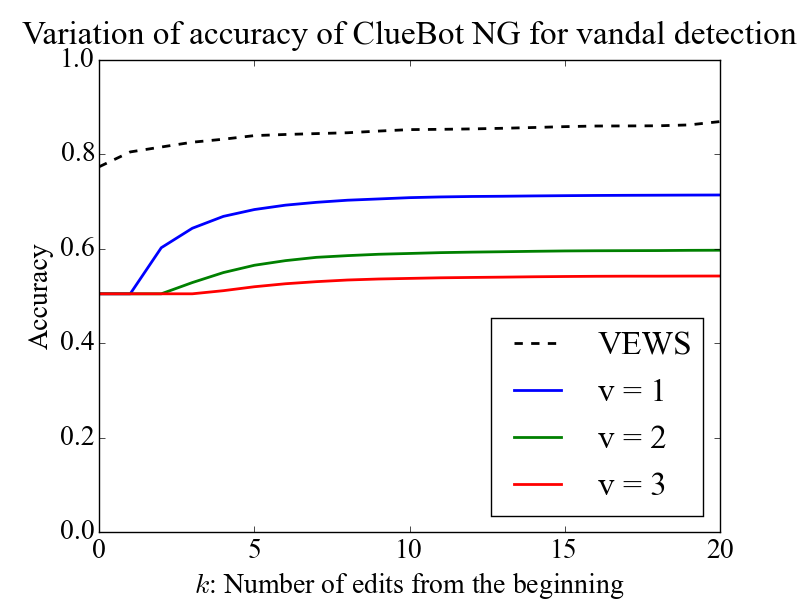}
  \caption{\small Plot showing the variation of accuracy for vandal detection by considering reversions made by ClueBot NG.}\label{fig:cluebot}
  \vspace*{-2mm}
\end{figure}

\emph{Comparison with ClueBot NG.} Given an edit, 
ClueBot NG \cite{cluebot} detects and reverts vandalism automatically. 
We could use ClueBot NG to classify a user as a vandal if he has made at least
 $v$ vandalism edits (edits that were reverted by ClueBot NG).
For comparing this heuristic with \VEWS\, we use $v=1,2,3$. Figure~\ref{fig:cluebot} shows that the maximum accuracy achieved by ClueBot NG is 71.4\% (when $v$ = 1) and accuracy decreases as $v$ increases. Therefore, \VEWS\ outperforms this use of ClueBot NG.

\emph{When does \VEWS\ Detect Vandals?}
Of 17,027 vandals in our dataset, \VEWS\ detects 3,746 that ClueBot NG does not detect (i.e. where ClueBot NG does not revert any edits by this person). In addition, it detects 7,795 vandals before ClueBot NG -- on average 2.6 edits before ClueBot NG did. In 210 cases, ClueBot NG detects a vandal edit 5.29 edits earlier (on average) than \VEWS\ detects the vandal and there are 1,119  vandals that ClueBot NG detects but \VEWS\ does not. Overall, when both detect the vandal, \VEWS\ does it 2.39 edits (on average) before ClueBot NG does.

Instead of reverts made by ClueBot NG, when we consider reverts made by any human or any known vandalism detection system, \VEWS\ detects the vandal at least as early as its first reversion in 87.36\% cases --- in 43.68\% of cases, \VEWS\ detects the vandal 2.03 edits before the first reversion. \emph{Thus, on aggregate, \VEWS\ outperforms both humans and other vandalism detection system in early detection of vandals}, though there are definitely a small number of cases (7.8\%) on which ClueBot NG performs very well.\footnote{\small We do not compare with STiki in this experiment, as it does not automatically revert edits.}

\emph{Comparison with STiki.} 
STiki provides a ``probability of vandalism'' score to each edit. STiki also maintains a user reputation score, which is developed by looking at the user's past edits (the higher the score, the higher the probability that the user is a vandal). We use both these scores separately to compare against STiki.

\begin{figure}[t]
\centering
  \includegraphics[trim = 1.4cm 0mm 0.6cm 0mm, clip, width=7.5cm]{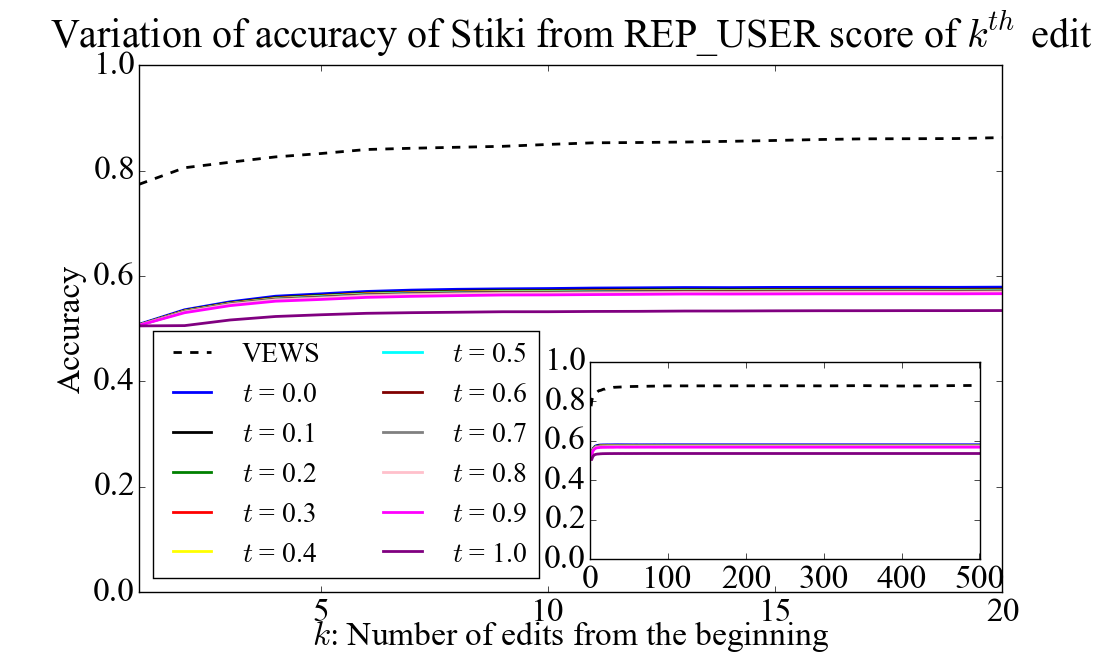}
  \caption{\small Plot showing the variation of accuracy for vandal detection by considering $k^{th}$ REP\_USER score given by STiki.}\label{fig:stiki_user}
  \vspace{-4mm}
\end{figure}

We first consider a user to be a vandal if his STiki reputation score (REP\_USER) after making the $k^{th}$ edit is at least $t$. Figure~\ref{fig:stiki_user} shows the results of this experiment where we vary $t$ from 0 to 1 in steps of 0.1. We also report the \VEWS\ curve for comparison.
We see that the STiki user reputation score to detect vandals has less than 60\% accuracy and is handily beaten by \VEWS. We do not test for values of $t$ greater than $1$ as accuracy decreases as $t$ increases.  

\begin{figure}[t]
\centering
  \includegraphics[trim =1.4cm 0cm 1cm 0.5cm, clip, width=7.8cm]{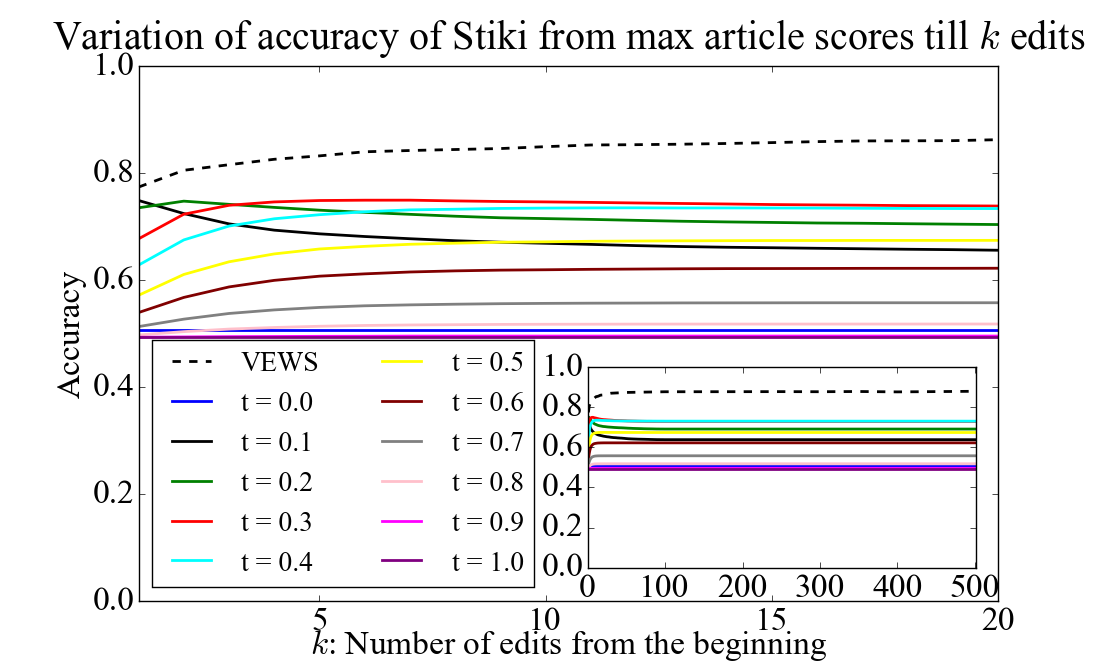}
  \caption{\small Plot showing the variation of accuracy for vandal detection by considering article scores given by STiki. RULE: If the user makes 1 edit in first $k$ that gets score $>$ $t$, then the user is a vandal.}\label{fig:stiki_edit1}
  \vspace{-5mm}
\end{figure}

In the second experiment, we say that a user is a vandal after making $k$ edits if the maximum STiki score among these $k$ edits is more than a threshold $t$.\footnote{\small We also tested using average instead of maximum with similar results.} We vary the values of $t$ from 0 to 1 and the results can be seen in Figure~\ref{fig:stiki_edit1}. We also did experiments for the case when we classify a user as a vandal if the maximum two and three scores are above $t$, which yielded lower accuracy scores.

\begin{figure}[t]
\centering
  \includegraphics[trim = 1.2cm 0cm 1.7cm 0.75cm, clip, width=5.5cm]{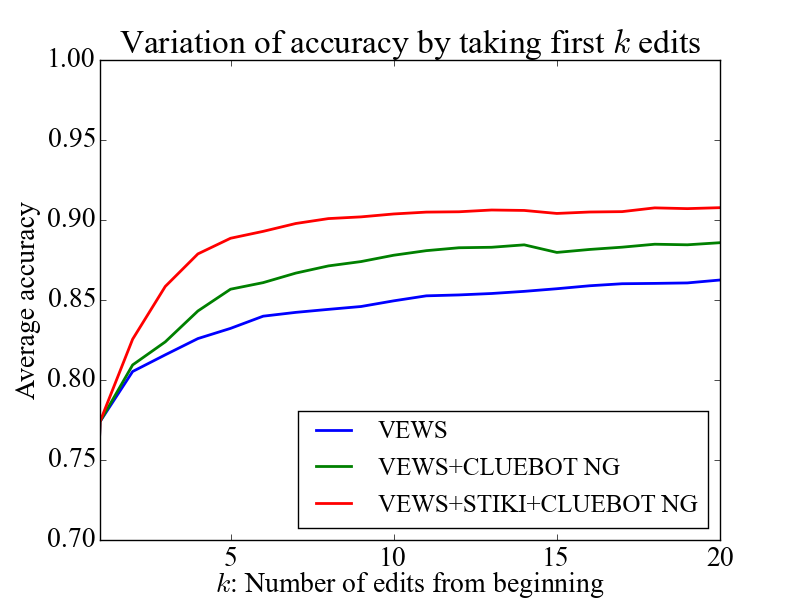}
  \caption{\small Figure showing effect of adding STiki and ClueBot NG's features to our \VEWS\ features.}\label{fig:all_max}
  \vspace{-4mm}
\end{figure}

\emph{Combining \VEWS, Cluebot NG and STiki.} 
\VEWS\ can be improved by adding linguistic and meta-data features from ClueBot NG and STiki.  In addition to the features in \VEWS, we add the following features: i) number of edits reverted\footnote{\small We allow these reverts to be considered as they are generated with no human input, so the resulting combination is still fully automated.} by ClueBot NG until the $k^{th}$ edit, ii) user reputation score by STiki after the $k^{th}$ edit, and iii) maximum article edit score given by STiki until the $k^{th}$ edit (we also did experiments with the average article edit score instead of maximum, which gave similar results).  Figure~\ref{fig:all_max} shows the variation of average accuracy by using the first-$k$ edits made by the user to identify it as a vandal.
The accuracy of the \VEWS-ClueBot combination is 88.6\% ($k=20$), which is higher than either of them alone. 
Observe that this combination does not consider any human input.
The accuracy of the combination \VEWS-ClueBot-STiki improves slightly to 90.8\% ($k=20$), but STiki considers human inputs while calculating its scores.

\vspace{-4mm}
\section{Conclusions}
In this paper, we develop a theory based on edit-pairs and edit-patterns to study the behavior of vandals on Wikipedia and distinguish these behaviors from those of benign users. We make the following contributions.

1. First, we develop the \raw\ dataset which contains a host of information about Wikipedia users and their behaviors. 

2. Second, we conduct a detailed analysis of behaviors that distinguish vandals from benign users. Notable distinctions that do not involve revert information include:

(a) We find that the first page edited by vandals is much more likely to be a normal page -- in contrast, benign users' first edits are much more likely to occur on meta-pages.

(b) We find that benign users take longer to edit a page than a vandal user. 

(c) We find that benign users are much more likely to re-edit the same page quickly (within 3 minutes) as compared to vandals, possibly because they wanted to go back and improve or fix something they previously wrote.

These are just three major factors that allow us to differentiate between vandals and benign users. Many others are detailed in the paper providing some of the first behavioral insights that do not depend on reverts that differentiate between vandals and benign users.

3. We develop three approaches to predict which users are vandals. Each of these approaches uses SVM with different sets of features. Our \VEWS\ algorithm provides the best performance, achieving 87.82\% accuracy. If in addition we consider temporal factors, namely that vandals next month are more likely to behave like vandals in the last few months, this accuracy goes up to 89.5\%. Moreover, we show that the combination of \VEWS\ and past work (ClueBot NG and STiki) increases accuracy to 90.8\%, even without any human generated reversion information. Moreover, \VEWS\ detects far more vandals than ClueBot NG. When both \VEWS\ and ClueBot NG predict vandals, \VEWS\ does it 2.39 edits (on average) before ClueBot NG does.

\smallskip
\textbf{Acknowledgements.} Parts of this work were supported by ARO grants W911NF11103, W911NF1410358, and\\W911NF09102.

\bibliographystyle{IEEEtran}
\vspace{-2mm}
\bibliography{www_bib}

\begin{thebibliography}{10}
\providecommand{\url}[1]{#1}
\csname url@samestyle\endcsname
\providecommand{\newblock}{\relax}
\providecommand{\bibinfo}[2]{#2}
\providecommand{\BIBentrySTDinterwordspacing}{\spaceskip=0pt\relax}
\providecommand{\BIBentryALTinterwordstretchfactor}{4}
\providecommand{\BIBentryALTinterwordspacing}{\spaceskip=\fontdimen2\font plus
\BIBentryALTinterwordstretchfactor\fontdimen3\font minus
  \fontdimen4\font\relax}
\providecommand{\BIBforeignlanguage}[2]{{%
\expandafter\ifx\csname l@#1\endcsname\relax
\typeout{** WARNING: IEEEtran.bst: No hyphenation pattern has been}%
\typeout{** loaded for the language `#1'. Using the pattern for}%
\typeout{** the default language instead.}%
\else
\language=\csname l@#1\endcsname
\fi
#2}}
\providecommand{\BIBdecl}{\relax}
\BIBdecl

\bibitem{vandalism}
\url{http://en.wikipedia.org/wiki/Wikipedia:Vandalism}.

\bibitem{openstreetmap}
P.~Neis, M.~Goetz, and A.~Zipf, ``Towards automatic vandalism detection in
  openstreetmap,'' \emph{ISPRS International Journal of Geo-Information},
  vol.~1, no.~3, pp. 315--332, 2012.

\bibitem{cluebot}
\url{http://en.wikipedia.org/wiki/User:ClueBot_NG}.

\bibitem{stiki}
http://en.wikipedia.org/wiki/Wikipedia:STiki.

\bibitem{snuggle}
\url{http://en.wikipedia.org/wiki/Wikipedia:Snuggle}.

\bibitem{Velasco10}
S.~M. Mola-Velasco, ``Wikipedia vandalism detection through machine learning:
  Feature review and new proposals - lab report for pan at clef 2010.'' in
  \emph{CLEF}, 2010.

\bibitem{WestKL10}
A.~G. West, S.~Kannan, and I.~Lee, ``Detecting wikipedia vandalism via
  spatio-temporal analysis of revision metadata?'' in \emph{EUROSEC}, 2010, pp.
  22--28.

\bibitem{AdlerAP10}
B.~T. Adler, L.~de~Alfaro, and I.~Pye, ``Detecting wikipedia vandalism using
  wikitrust - lab report for {PAN} at {CLEF} 2010,'' in \emph{{CLEF}}, 2010.

\bibitem{AdlerAMRW11}
B.~T. Adler, L.~de~Alfaro, S.~M. Mola{-}Velasco, P.~Rosso, and A.~G. West,
  ``Wikipedia vandalism detection: Combining natural language, metadata, and
  reputation features,'' in \emph{CICLing}, 2011, pp. 277--288.

\bibitem{Potthast}
M.~Potthast, B.~Stein, and R.~Gerling, ``Automatic vandalism detection in
  wikipedia,'' in \emph{ECIR}, 2008, pp. 663--668.

\bibitem{missing}
M.~Sumbana, M.~Gonçalves, R.~Silva, J.~Almeida, and A.~Veloso, ``Automatic
  vandalism detection in wikipedia with active associative classification,'' in
  \emph{TPDL}, 2012, pp. 138--143.

\bibitem{ferschke2013survey}
O.~Ferschke, J.~Daxenberger, and I.~Gurevych, ``A survey of nlp methods and
  resources for analyzing the collaborative writing process in wikipedia,'' in
  \emph{The People's Web Meets NLP}.\hskip 1em plus 0.5em minus 0.4em\relax
  Springer, 2013, pp. 121--160.

\bibitem{wikitrust}
\url{http://www.wikitrust.net/}.

\bibitem{WestL12}
R.~West and J.~Leskovec, ``Human wayfinding in information networks,'' in
  \emph{WWW}, 2012, pp. 619--628.

\bibitem{cockburn2001web}
A.~Cockburn and B.~McKenzie, ``What do web users do? an empirical analysis of
  web use,'' \emph{International Journal of human-computer studies}, vol.~54,
  no.~6, pp. 903--922, 2001.

\bibitem{catledge1995characterizing}
L.~D. Catledge and J.~E. Pitkow, ``Characterizing browsing strategies in the
  world-wide web,'' \emph{Computer Networks and ISDN systems}, vol.~27, no.~6,
  pp. 1065--1073, 1995.

\bibitem{adar}
E.~Adar, J.~Teevan, and S.~T. Dumais, ``Large scale analysis of web
  revisitation patterns,'' in \emph{SIGCHI}.\hskip 1em plus 0.5em minus
  0.4em\relax ACM, 2008, pp. 1197--1206.

\bibitem{Welser11}
H.~T. Welser, D.~Cosley, G.~Kossinets, A.~Lin, F.~Dokshin, G.~Gay, and
  M.~Smith, ``Finding social roles in wikipedia,'' in \emph{iConference}, 2011,
  pp. 122--129.

\bibitem{wikiApi}
\url{https://www.mediawiki.org/wiki/API}.

\bibitem{hopsDB}
\url{http://beta.degreesofwikipedia.com/}.

\bibitem{revertDB}
A.~Halfaker, \url{http://datahub.io/dataset/english-wikipedia-reverts}.

\bibitem{Kittur}
A.~Kittur, B.~Suh, B.~A. Pendleton, and E.~H. Chi, ``He says, she says:
  Conflict and coordination in wikipedia,'' in \emph{SIGCHI}, 2007, pp.
  453--462.

\bibitem{stikiAPI}
\url{http://armstrong.cis.upenn.edu/stiki_api.php?}

\bibitem{geurts2006extremely}
P.~Geurts, D.~Ernst, and L.~Wehenkel, ``Extremely randomized trees,''
  \emph{Machine learning}, vol.~63, no.~1, pp. 3--42, 2006.

\bibitem{autoencoder}
Y.~Bengio, ``Learning deep architectures for ai,'' \emph{Foundations and
  trends{\textregistered} in Machine Learning}, vol.~2, no.~1, pp. 1--127,
  2009.

\bibitem{scikit}
F.~Pedregosa, G.~Varoquaux, A.~Gramfort, V.~Michel, B.~Thirion, O.~Grisel,
  M.~Blondel, P.~Prettenhofer, R.~Weiss, V.~Dubourg, J.~Vanderplas, A.~Passos,
  D.~Cournapeau, M.~Brucher, M.~Perrot, and E.~Duchesnay, ``Scikit-learn:
  Machine learning in {P}ython,'' \emph{Journal of Machine Learning Research},
  vol.~12, pp. 2825--2830, 2011.

\bibitem{theano}
J.~Bergstra, O.~Breuleux, F.~Bastien, P.~Lamblin, R.~Pascanu, G.~Desjardins,
  J.~Turian, D.~Warde-Farley, and Y.~Bengio, ``Theano: a {CPU} and {GPU} math
  expression compiler,'' in \emph{SciPy}, Jun. 2010.

\end{thebibliography}

\end{document}